\documentclass[
reprint,
superscriptaddress,
showpacs,
preprintnumbers,
nofootinbib,
nobibnotes,
amsmath,
amssymb, 
aps,
prd,
dvipsnames,
floatfix
]{revtex4-1}

\usepackage[utf8]{inputenc}
\usepackage[normalem]{ulem}
\usepackage{graphicx}
\usepackage{dcolumn}
\usepackage{url}
\usepackage{enumerate}

\usepackage{slashed,multirow,relsize,soul,feynmp-auto,tikz}
\usepackage{color}
\usepackage{mathrsfs} 
\usepackage{amsmath}
\usepackage{cancel}
\usepackage{bbold}
 \usepackage{mathrsfs}
\usepackage{braket}
\usepackage{physics}
\usepackage{multirow}
\usepackage{xspace}
\usepackage{xcolor}

\usepackage[colorlinks=true,allcolors=purple]{hyperref}
\usepackage[capitalize]{cleveref}

\usepackage{fontawesome} 
\definecolor{blue-violet}{rgb}{0.33, 0.17, 0.89}

\newcommand {\be}{\begin{equation}}
\newcommand {\ee}{\end{equation}}
\newcommand {\ba}{\begin{eqnarray}}
\newcommand {\ea}{\end{eqnarray}}

\def\e6{E(6)}
\def\10{SO(10)}
\def\21{SA(2) $\otimes$ U(1) }
\def\321{$\mathrm{SU(3) \otimes SU(2) \otimes U(1)}$ }

\def\422{SA(4) $\otimes$ SA(2) $\otimes$ SA(2)}

\def\roughly#1{\mathrel{\raise.3ex\hbox{$#1$\kern-.75em
      \lower1ex\hbox{$\sim$}}}} \def\lsim{\roughly<}
\def\gsim{\roughly>}

\def\lsim{\raise0.3ex\hbox{$\;<$\kern-0.75em\raise-1.1ex\hbox{$\sim\;$}}}
\def\gsim{\raise0.3ex\hbox{$\;>$\kern-0.75em\raise-1.1ex\hbox{$\sim\;$}}}

\newcommand{\Tehran}{%
	School of physics, Institute for Research in Fundamental Sciences (IPM)
	\\
	P.O.Box 19395-5531, Tehran, Iran}
\def\roughly#1{\mathrel{\raise.3ex\hbox{$#1$\kern-.75em
			\lower1ex\hbox{$\sim$}}}} \def\lsim{\roughly<}
\def\gsim{\roughly>}

\def\lsim{\raise0.3ex\hbox{$\;<$\kern-0.75em\raise-1.1ex\hbox{$\sim\;$}}}
\def\gsim{\raise0.3ex\hbox{$\;>$\kern-0.75em\raise-1.1ex\hbox{$\sim\;$}}}

\definecolor{palatinate}{rgb}{0.494, 0.192, 0.482}

\newcommand{\kmevent}{KM3-230213A\xspace}

\begin{document}

\title{Astrophysical flux of dark particles as a solution to the KM3NeT and IceCube tension over KM3-230213A}

\author{Yasaman Farzan}
\email{yasaman@theory.ipm.ac.ir}
\affiliation{\Tehran}
\author{Matheus Hostert}
\email{mhostert@g.harvard.edu}
\affiliation{Department of Physics \& Laboratory for Particle Physics and Cosmology, Harvard University, Cambridge, MA 02138, USA}

\date{\today}

\begin{abstract}
We entertain the possibility that transient astrophysical sources can produce a flux of dark particles that induce ultra-high-energy signatures at neutrino telescopes such as IceCube and KM3NeT.
We construct scenarios where such ``dark flux" can produce meta-stable dark particles inside the Earth that subsequently decay to muons, inducing through-going tracks in large-volume neutrino detectors.
We consider such a scenario in light of the $\mathcal{O}(70)$~PeV ultra-high-energy muon observed by KM3NeT and argue that because of its location in the sky and the strong geometrical dependence of the signal, such events would not necessarily have been observed by IceCube.
Our model relies on the upscattering of a new particle $X$ onto new metastable particles that decay to dimuons with decay lengths of $\mathcal{O}(100)$~km.
This scenario can explain the observation by KM3NeT without being in conflict with the IceCube data.
\end{abstract}

\maketitle


\section{Introduction}

On the 13th of February of 2023, one of the two detectors of the KM3NeT project, Astroparticle Research with Cosmics in the Abyss (ARCA), observed an ultra-high energy (UHE) through-going muon pointing back to just above the horizon~\cite{KM3NeT:2025npi}.
The muon was reconstructed with an energy of $70$~PeV, with a $90\%$ confidence interval of $35 \text{ PeV} < E_\mu <380\text{ PeV}$.
The top of the ARCA detector is located at a depth of $\sim 2.8 - 3.5$~km under the sea off the coast of Sicily in the Mediterranean sea.
Because of the significant amount of overburden, muons from this direction and of such energies are extremely unlikely to be of atmospheric origin.
Instead, they can be produced from charged-current interactions of neutrinos in the water or rock leading up to the detector.
Under this interpretation, the $\nu_\mu$ or $\bar{\nu}_\mu$ that produced the muon should have a median energy of 220~PeV with a 90\% confidence interval of $72~\text{PeV} < E_\nu < 2.6~\text{EeV}$~\cite{KM3NeT:2025npi}. 
This is the leading hypothesis behind such an event, although the origin of such a neutrino remains a mystery. 
At such high energies, the atmospheric neutrino flux is too small to lead to such an event in KM3NeT, favoring an astrophysical interpretation for \kmevent.
The direction of the incoming muon is determined at KM3NeT with a $1.5^\circ$ angular precision, currently limited by the systematic uncertainty on the absolute orientation of the detector.
While a few candidate sources were identified within this region of the sky, none were observed in coincident multi-messenger signals, such as with gamma rays~\cite{KM3NeT:2025aps,KM3NeT:2025bxl}.

\begin{figure*}[t]
    \centering
    \includegraphics[width=\textwidth]{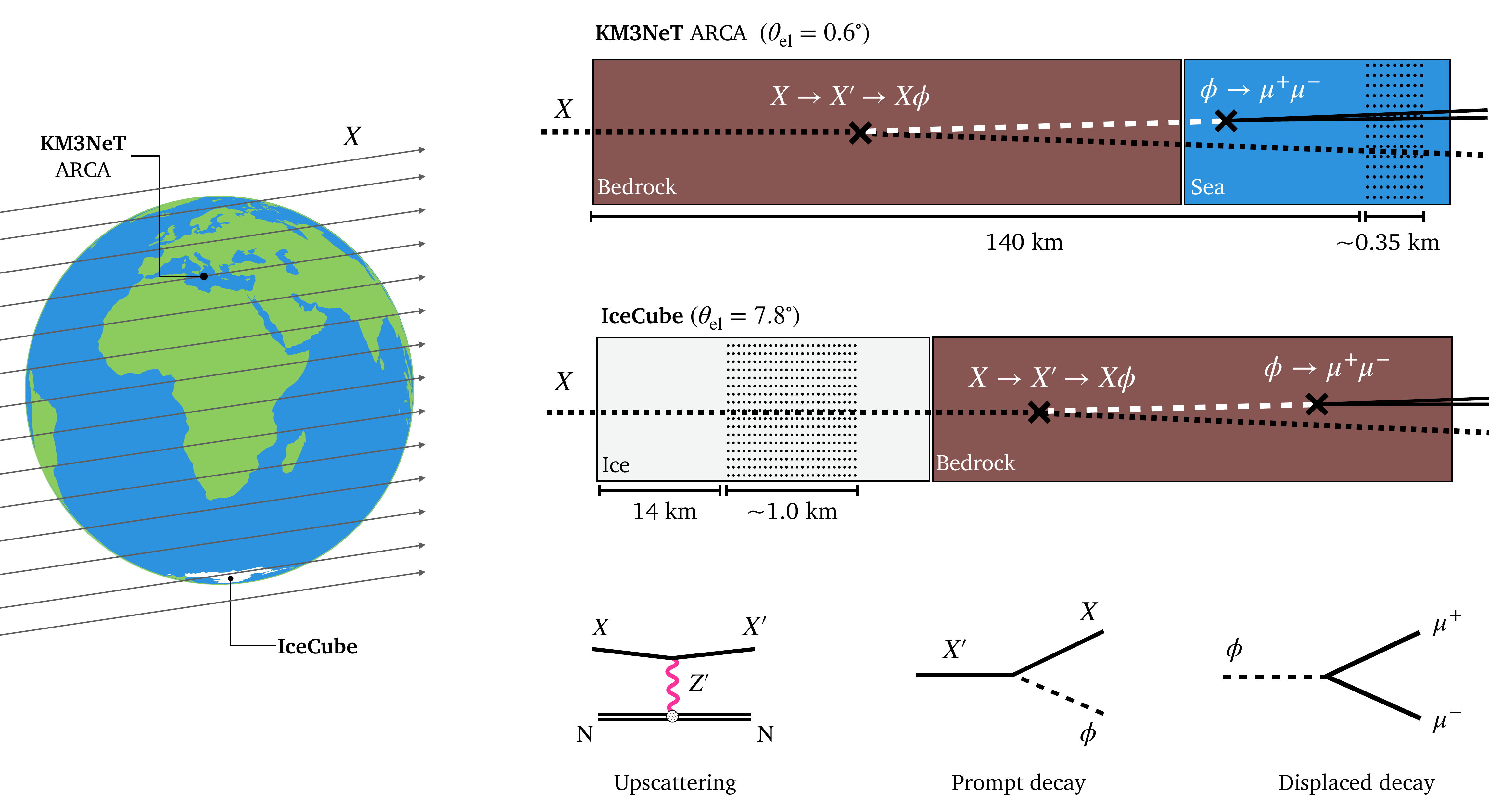}
    \caption{A schematic  one-dimensional view of the KM3NeT and IceCube setups and the event topology (not to scale). 
    A transient astrophysical flux of $X$ travels from left to right, upscattering to $X'$ within the Earth. The produced $X'$ promptly decays back to $X$ and a metastable new scalar, $\phi$.  The subsequent decay of $\phi$  to dimuons in the vicinity of KM3NeT produces the observed signal.
    The column depth before IceCube is much smaller than that before KM3NeT in the direction of \kmevent.
    }
    \label{fig:sketch}
\end{figure*}

If \kmevent is indeed an astrophysical neutrino, the astrophysical source behind it should also have been observed by the IceCube neutrino telescope~\cite{IceCube:2016zyt}.
IceCube has a significantly longer exposure time, with over a decade of operation to be compared with less than a year for ARCA, and its instrumented volume is also much larger, spanning a volume closer to $1$~km$^3$, to be compared with the $0.15$~km$^3$ volume of ARCA.
Despite that, IceCube has not detected any note-worthy tracks in the same direction as \kmevent. 
The statistical tension between these two experiments was analyzed in detail in Ref.~\cite{Li:2025tqf} (see also \cite{KM3NeT:2025ccp}), which concluded that even in the most favorable case of a transient point source, the tension between the non-observation of such a neutrino source by IceCube and the observation of \kmevent by KM3NeT is greater than 2$\sigma$ CL.
For diffuse sources, such as cosmogenic neutrinos from cosmic-ray interactions in the intergalactic medium~\cite{KM3NeT:2025vut}, the tension is much larger, between $3.1-3.6\sigma$~\cite{Li:2025tqf}.
While many interpretations have been discussed in the context of the KM3NeT observation~\cite{Muzio:2025gbr,Wu:2024uxa,Boccia:2025hpm,Kohri:2025bsn,Borah:2025igh,Narita:2025udw,Alves:2025xul,Choi:2025hqt,Zantedeschi:2024ram,Airoldi:2025opo}, only a few have attempted to address the tension~\cite{Brdar:2025azm}.

Motivated by this puzzle, we revisit the KM3NeT event in the context of new physics and show that if the muon is produced as a result of  the scatterings of a transient astrophysical  flux of dark particles, the tension between the two experiments can be entirely resolved.
As pointed out in \cite{Brdar:2025azm}, for a fixed location in the sky, the distance between the detector and the entry point on the surface of the Earth is $L_{\rm KM}\sim 140$~km in the case of KM3NeT but only $L_{\rm IC}\sim14$ km for IceCube. 
Ref.~\cite{Brdar:2025azm} argued that if the event were caused by a UHE astrophysical sterile neutrino that resonantly oscillates to $\nu_\mu$ due to matter effects, then IceCube would not necessarily have observed it due its much smaller Earth column depth in the direction of the source.

In this paper, we use a similar approach but argue that the muon can be induced instead by the scattering and decay of new dark particles propagating through the Earth.
Our scenario relies on a flux of UHE astrophysical dark particles beyond the Standard Model (SM) that interact with matter in the Earth to produce new metastable particles that subsequently decay to muons in the vicinity of KM3NeT.
In these scenarios, the visible through-going track signature is produced by a collimated $\mu^+\mu^-$ pair.
Since the \kmevent track has  propagated only $\mathcal{O}(300 - 400)$~m through the instrumented volume, we argue that the   stochastic energy loss by the dimuon pairs is still sufficient to provide an acceptable explanation to the $dE/dX$ profile of the event.\footnote{In fact, taus could also produce through-going tracks at these energies, however, due to the larger tau mass and its smaller energy loss, such an interpretation would require an even larger lepton energy to have triggered all 3,672 PMT hits in KM3NeT.}
We note that the upscattering of dark particles in the Earth also appears in many new-physics explanations of the anomalous events at ANITA-IV~\cite{Esmaili:2019pcy,Bertolez-Martinez:2023scp} (see also similar discussions in~\cite{BertolezMartinez2025}), although the requirements for the duration of the source, the scattering and decay lengths, as well as the visible products of the new particles are very different in our scenario.

The rest of this article is divided as follows.
In \cref{sec:experiments}, we discuss the ARCA-21 and IceCube detectors in greater detail in the context of \kmevent.
In \cref{sec:darkbursts}, we present our new physics mechanism to resolve their tension and calculate the cross sections and lifetimes 
of the new particles necessary to realize it.
We speculate on the potential sources of dark particles in \cref{sec:sources} and present concrete particle-physics models that realize our idea in \cref{sec:models}. We also briefly discuss the bounds from cosmology and the terrestrial experiments on our model. We then discuss the prospect to discover the new particles of our model in the upcoming experiments.
We summarize and present our conclusions in \cref{sec:conclusions}.
\section{ARCA-21 and IceCube: \kmevent}
\label{sec:experiments}

\begin{figure*}[t]
    \centering
    \includegraphics[width=\textwidth]{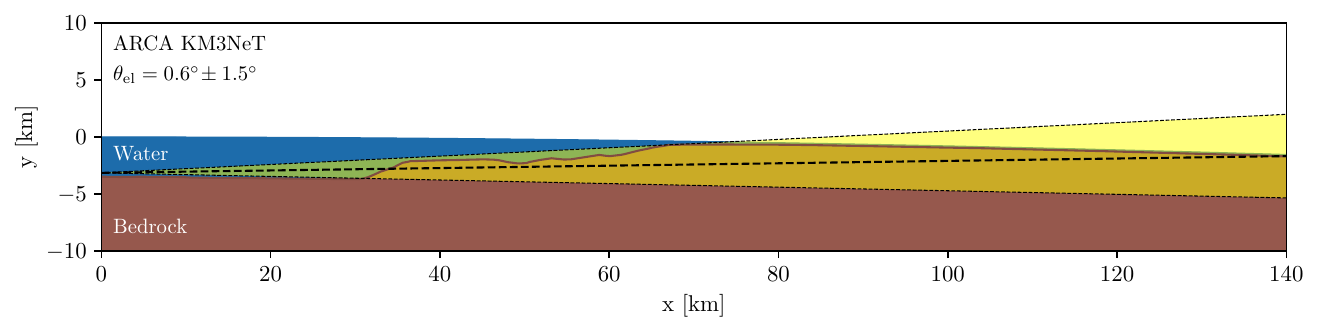}
    \includegraphics[width=\textwidth]{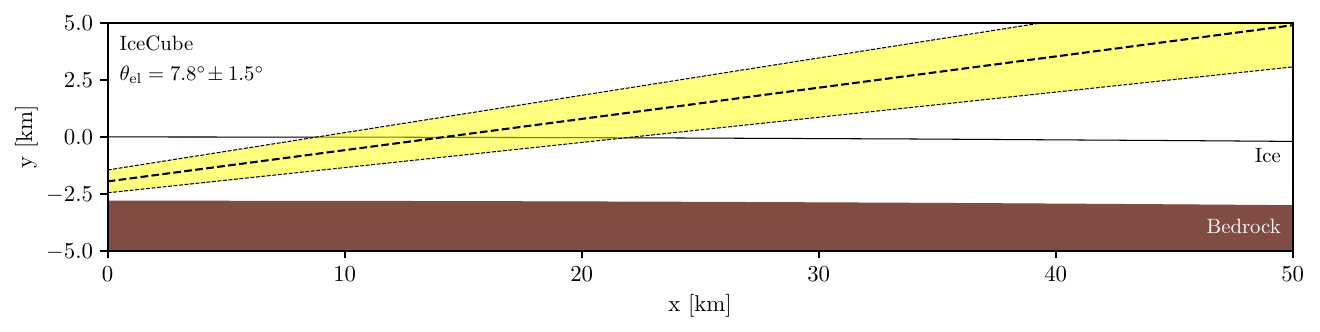}
    \caption{An illustration of the incoming direction of the $X$ particles for \kmevent at ARCA (top) and IceCube (bottom) to-scale.
    The yellow shaded region shows the $1\sigma$ uncertainty on the line of sight to the source.
    The $X$ particles would travel from right to left (West to East).
    \label{fig:diagram}}
\end{figure*}

KM3NeT consists of two detectors deep under the Mediterranean sea: ARCA, located off the coast of Capo Passero, Sicily, and optimized for high-energy neutrino astronomy; and ORCA, situated off Toulon, France, and designed for studying the oscillations of atmospheric neutrinos at lower energies. 
The UHE event was observed with ARCA-21, an early configuration of the ARCA detector that operated with 21 vertical strings of $700$~m in height, instrumenting a cylindrical volume of approximately 0.15~km$^3$~\cite{Biagi:2023yoy}.
This is only a small fraction of the total volume of about 1~km$^3$ that ARCA will occupy once it is completed~\cite{KM3NeT:2018wnd}.
IceCube is a similar detector located deep in the ice in the South Pole~\cite{IceCube:2016zyt}.
IceCube, in comparison, has operated with a total of 86 strings in the ice since 2011, spanning a volume of approximately $1$~km$^3$. 
The total live time for the ARCA-21 dataset amounts to 287.4 days, which is, again, much shorter than IceCube’s dataset that spans over 14 years of operation.
These differences are the primary cause for the tension between the two experiments when it comes to interpretations of \kmevent in diffuse or steady astrophysical neutrino source models.

Both ARCA and IceCube are designed to detect ultra-high-energy (UHE) astrophysical neutrinos by observing Cherenkov light emitted by secondary charged particles produced in neutral-current (NC) and charged-current (CC) neutrino interactions with matter.
Cherenkov light is detected with arrays of Digital Optical Modules (DOMs) deployed on the vertical strings. 
Electrons and hadrons typically generate localized showers called cascades, while muons and highly-boosted tau leptons leave extended track signatures. 
At PeV energies, through-going tracks, like muons that enter and exit the detector, deposit only a small fraction of their energy within the instrumented volume but enable superior angular reconstruction due to their long, straight paths.
\kmevent is a through-going track event, meaning its arrival direction is tightly constrained, limited primarily by systematic uncertainties in the detector’s absolute orientation. 
However, the energy of the muon is highly uncertain given that its starting position is not known and that the muon propagation range at these energies is as large as $\mathcal{O}(20-40)$~km~\cite{Koehne:2013gpa,Garg:2022ugd,Cummings:2023iuw}.

The reconstructed equatorial coordinates of \kmevent are a right ascension of $\alpha = 94.3^\circ$, declination of $\delta = -7.8^\circ$, and a detection time of $\text{MJD}=59988.0533299$~\cite{KM3NeT:2025npi}. 
The angular uncertainty on the reconstructed track direction is $1.5^\circ$ ($2.2^\circ$) at 68\% (90\%) confidence level.
At ARCA’s location, these celestial coordinates correspond to a local elevation angle of approximately $\theta_{\rm el}^{\rm KM} = 0.6^\circ$ and an azimuth angle of $\phi^{\rm KM}=259.8^\circ$, indicating a nearly-horizontal event traveling Eastwards.
Translating these celestial coordinates to local IceCube coordinates, we find a local elevation angle of $\theta_{\rm el}^{\rm IC} = 7.8^\circ$ and an azimuth angle of $\phi^{\rm IC} = 292.5^\circ$.
Therefore, at IceCube, the incoming flux of neutrinos or new particles would be significantly more down-going than at ARCA, although still comfortably within IceCube's field of view.
For neutrinos, the horizon direction is optimal as for very negative elevation angles, the Earth becomes opaque while for very large and positive elevation angles the probability for the neutrino to interact before the detector is too small and backgrounds too large.

\cref{fig:diagram} shows a to-scale view of the ARCA and IceCube geometries in the context of \kmevent.
Due to the angular uncertainty, the column depth $X_{\rm depth} = \int \dd x \rho(x)$ seen by the neutrino or dark particle $X$ on its way to ARCA can significantly change  depending on the exact location of the source in the sky.
Given the direction of the event, the uncertainty on the elevation angle, and not the azimuth, is the most relevant.
By extrapolating \emph{Extended Data Fig.~4} in Ref.~\cite{KM3NeT:2025npi} and varying $-0.9^\circ < \theta_{\rm el} < 2.1^\circ$, we estimate the range of distances to be
\begin{align}
    74\text{ km} &< L_{\rm KM} < 320\text{ km},
    \\\nonumber
   7.4\text{ t/cm}^2 &< X_{\rm depth}^{\rm KM} < 78\text{ t/cm}^2.
\end{align}
This estimate assumes the seabed rock density to be constant and equal to $2.6$~g/cm$^3$.
At IceCube, the corresponding range, including the height of the detector in this case, is
\begin{align}
    8.7\text{ km} &< L_{\rm IC} < 21\text{ km},
    \\\nonumber
   0.8\text{ t/cm}^2 &< X_{\rm depth}^{\rm IC} < 2.0\text{ t/cm}^2.
\end{align}
Since the elevation angle at KM3NeT is not necessarily fully correlated with that of IceCube's, the ratio of the column depths traversed by neutrinos or new particles is within the following $68\%$ ($90\%$) confidence interval:
\begin{equation}
    1.0 \times 10^{-2} \,(6.7 \times 10^{-3}) <  \frac{X_{\rm depth}^{\rm IC} }{X_{\rm depth}^{\rm KM} } < 0.27 \,(0.38).
\end{equation}

\begin{figure}[t]
    \centering
    \includegraphics[width=0.49\textwidth]{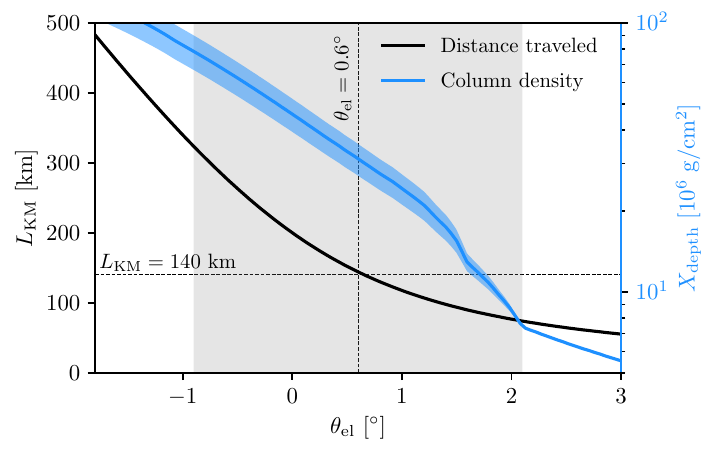}
    \caption{The distance traveled by the new particles through the Earth before reaching KM3NeT, $L_{\rm KM}$, as a function of the elevation angle of the source at the KM3NeT location. 
    Also shown in blue is the column depth $X_{\rm depth}$, with the band illustrating the uncertainty from the local rock density of 2.2~g/cm$^3<\rho_{\rm rock} < 3.0$~g/cm$^3$.
    \label{fig:elevations}}
\end{figure}

In \cref{fig:elevations}, we show the $\theta_{\rm el}$ dependence of the distance traveled and the column depth seen by the incoming particles before reaching the detector.
The band for the column depth shows the range of values for 2.2~g/cm$^3<\rho_{\rm rock} < 3.0$~g/cm$^3$.
At IceCube, because the event is coming from higher above the horizon and can only cross ice, the relative uncertainty on $L_{\rm IC}$ and $X_{\rm depth}^{\rm IC}\simeq \rho_{\rm ice} L_{\rm IC} \simeq 1.3$~t/cm$^2$ is far smaller.

\section{Dark Bursts at KM3NeT and IceCube}
\label{sec:darkbursts}

In this section, we elaborate on how bursts of dark particles can explain the \kmevent event without predicting corresponding signatures in IceCube.
We introduce the mechanism below, discuss the upscattering process in \cref{sub:scattering}.  We then  describe the properties of the metastable particles and its decay to dimuon in \cref{sub:Metastable}.
In \cref{sub:muondetection}, we present formulas for the number of UHE muons at KM3NeT and compare it to that at IceCube.  
In \cref{sub:duration}, we discuss the  duration and magnitude of the dark flux.

In each experiment, a plane-wave flux of $X$ particles passes mostly unimpeded through a column depth of rock, ice, or water before reaching the KM3NeT or IceCube detector.
If a muon is produced along the way, it can be detected once it crosses the instrumented region provided it has enough energy.
Consider, for instance, the following scattering in the Earth before $X$ reaches the KM3NeT detector,
\begin{align}\label{eq:scattering}
    X + \text{nucleus/electron} \to X' + \text{nucleus/electron},
\end{align}
where $X'$ is another dark particle that is heavier than $X$.

In our model, $X'$ promptly decays to $X$ and a new meta-stable particle, $\phi$, as shown in \cref{fig:sketch}. 
The subsequent decay of $\phi$ to dimuon explains the KM3-230213A signal, provided that the $\phi$ decay takes place in the vicinity of the detector (at a distance within the muon range from the detector).
Therefore, every upscattering is assumed to be followed by the following decay chain
\begin{equation}
    \text{Prompt: } X' \to X +\phi,  \quad \text{Displaced: }\phi \to \mu^+ \mu^-.
\end{equation}
In IceCube, the same process can also happen, albeit with a much smaller probability due to the smaller column depth.
In this construction, the tension between the two experiments is then alleviated by one main factor: not as many $X'$ particles are produced in the smaller Earth column depth of IceCube.

The energy loss of UHE muons is dominated by stochastic radiative processes like photo-nuclear and bremsstrahlung energy loss.
For through-going tracks, this implies that single muons should display instances of catastrophic energy loss, effectively inducing small cascades along the track-like signature.
The presence of three such distinctive cascades in \kmevent indicates that there is some stochasticity to the energy loss and helps identify it as a through-going muon.
However, given the poor energy resolution to tracks and that the stochasticity is only sampled in a relatively small volume with a relatively sparse DOM grid, we argue that such behavior can be easily mimicked by a muon pair.
While it is true that the energy loss profile of a collimated muon pair is more continuous than that of a single muon, for \kmevent, it is unlikely that the experiment would be able to tell the difference.

When the $X$ particles scatter off nuclei or off the electrons, they can in principle induce a shower or cascade like event. 
Let us estimate the number of such events at IceCube. 
Note that only a fraction $f_X$ of the $X$ particles that scatter on its way to KM3NeT produce an observable UHE through-going track.
That is, for KM3NeT to have observed a single UHE event, a total number of $f_X^{-1}$ of $X \to X'$ scattering events should have taken place during the time-window of the source at KM3NeT. 
The number of $X \to X'$ scattering per unit mass in the column depth of KM3NeT is approximately
\begin{equation}
\mathcal{C}\equiv
(f_X L_{\rm KM}  {A}_{\rm KM}\rho_{\rm rock})^{-1},
\end{equation}
where ${A}_{\rm KM}=0.364$ km$^2$ (see \cref{sub:muondetection}) and  $\rho_{\rm rock}\sim 2.6 \, {\rm g}/{\rm cm}^3$ is the average density of the seabed rock. 
At IceCube, the total number of scattering events inside the detector volume is then 
\begin{equation} \label{50-500}
\mathcal{C}{V_{\rm IC} \rho_{\rm ice}} \sim 0.7 \times\left( \frac{V_{\rm IC}}{1 \text{ km}^3}\right)\left(\frac{\rho_{\rm ice}}{0.92 \frac{\rm g}{\rm cm^3}}\right) \left(\frac{10^{-2}}{f_X}\right),
\end{equation}
which implies a  small number of scattering events inside IceCube. 
Moreover, as we shall see our model predicts the recoil energy of scattering to be below the detection threshold of IceCube.

\subsection{$X\to X'$ upscattering 
\label{sub:scattering}}

Let us consider generic couplings between two distinct dark particles $X$ and $X'$ with a light mediator $Z'$ as well as couplings to SM fermions of the form
\begin{align}\label{eq:generic_couplings}
    \mathcal{L}_{\rm scatt.} &= Z^\prime_\mu \left(g_e \bar{e}\gamma^\mu e+ \sum_{q=u}^t g_q \bar{q} \gamma^\mu q\right) 
    \\\nonumber & \qquad + g_X Z^\prime_\mu \left(\bar{X}^\prime \gamma^\mu X 
    + \text{ h.c.}\right)
\end{align}
The $Z'$ coupling to matter particles and its off-diagonal coupling between $X$ and $X'$ will lead to the upscattering process on nuclear and electron targets as in \cref{eq:scattering}.
Potential UV origin for these couplings are discussed in \cref{sec:models}.

Scattering via $t$-channel exchange of light vector boson peaks at the forward direction where the momentum exchange $Q^2 \equiv -t \ll s$, with $\sqrt s$ the center-of-mass energy of the reaction.  
Taking only the $Z'$ coupling to the hadronic vector current with coupling strength $g_N$, the cross section for $XN \to X' N$ on a fermionic  target can be written as
\begin{widetext}
\begin{align}\label{cross-s}
\frac{\dd\sigma_X}{\dd Q^2} = \frac{g_X^2 g_N^2 }{32 \pi m_N^2 |\vec k_X|^2} 
\left(
\frac{s_- s_-' + (s_--Q^2)(s_-'-Q^2)
- 2Q^2 m_X m_{X'} - 4 m_N^2 (s - m_X m_{X'}) + 2 m_N^4
}
{(Q^2+m_{Z'}^2)^2}
\right) |F(Q^2)|^2
\end{align}
\end{widetext}
where $s_-=s-m_{X}^2$ and $s_-^{'}=s-m_{X^{'}}^2$, $|\vec k_{X}|$ is the lab-frame momentum of the incoming $X$ particle, and $F(Q^2)$ is some form factor or structure function of the target $N$.
At UHE, where $s \gg Q^2$, the cross section simplifies to
\begin{equation}
    \frac{\dd \sigma_X}{\dd Q^2} \simeq \frac{g_X^2 g_N^2 }{4 \pi} 
\left[
\frac{1 + \frac{m_X^2-m_{X'}^2-Q^2}{2 E_X m_N}}
{(Q^2+m_{Z'}^2)^2}
\right] |F(Q^2)|^2. \label{simFQ}
\end{equation}

For coherent scattering on a nucleus (A,Z), the cross section is enhanced according to $g_N = Z g_p + N g_n$, where $g_p = 2 g_u + g_d$ and $g_n = g_u + 2g_n$.
The nuclear form factor $F(Q^2)$ then enforces that $Q \lesssim 1/R_{\rm nucleus} \sim \Lambda_{\rm QCD}/A^{1/3}$, where $A=Z+N$.
One can check that the coherence condition is easily satisfied even for relatively heavy dark particles $X$, as at UHEs, the minimum $Q^2$ of the reaction is tiny and approximately given by,
\begin{equation}
    Q^2_{\rm min} \simeq \frac{(M_{X'}^2 - M_X^2)^2}{4 E_X^2}.
\end{equation}

For intermediate values of $Q^2 \sim (0.02 \to 1)$ GeV$^2$, the scattering will be mostly (quasi-)elastic (QE) on nucleons.
In that case, we use \cref{cross-s} replacing $F(Q^2) \to F_{1}(Q^2)$, the Dirac form factor of the nucleon that falls off with a dipole mass of $M_D\simeq 0.84$~GeV.

For  $Q^2 \gg {\rm GeV}^2$, the scattering is in Deep Inelastic Scattering (DIS) regime. 
In that case, we can approximately describe the scattering on quarks by taking $M = m_q \simeq 0$, replacing $Q^2 \to Q^2 x$ and $F(Q^2) \to F(Q^2, x)\simeq f_q(x) + f_{\bar q}(x) \simeq 0.25 x^{-1.1}$ \cite{Gayler:2002ve}, and integrating over $x = Q^2/2 E_X M_N y$, with $y = E_{X'}/E_X$ being the inelasticity.
In our estimates, we take the $Z'$ couplings to the quarks to be universal and equal to $g_B =3 g_q$ for all $q = {u,d,s}$ and neglect the heavy quark content of the nucleon.
The end result for a light mediator is a  DIS cross section an order of magnitude smaller than the cross sections of the coherent or QE scattering. 
Thus, we are justified in neglecting the DIS regime.
Evaluating the integrals numerically, we find that all cross sections have a very weak energy dependence above $1$~PeV.
For instance, for $X$ upscattering on ${}^{28}$Si with $m_{Z'} = 300$~MeV, $m_{X} = 0$, and $m_{X'} = 5$~GeV, we find
\begin{align} \label{eq:coh-QE-DIS}
    \sigma_X^{\rm coh} \simeq 2.8\times 10^{-34} ~{\rm cm}^2\left(\frac{g_X g_B}{10^{-4}} \right)^2, 
    \\ 
    \sigma_X^{\rm QE} \simeq 5.7\times 10^{-35} ~{\rm cm}^2\left(\frac{g_X g_B}{10^{-4}} \right)^2, 
    \\ 
    \sigma_X^{\rm DIS} \simeq 7.2\times 10^{-36} ~{\rm cm}^2\left(\frac{g_X g_B}{10^{-4}} \right)^2.
\end{align}

The cross section of the scattering of $X$ off the electron is given by Eq. (\ref{simFQ}), replacing $g_N \to g_e$, $m_N\to m_e$ and $F(Q^2) \to 1$. 
The threshold for $X'$ production can be much larger in this case, 
\begin{equation}
    E_X^{\rm th} =  m_{X'} + \frac{m_{X’}^2 - m_X^2}{2 m_e} \simeq 25\text{ TeV},
\end{equation}
where we took $m_X = 0$ and $m_{X'} = 5$~GeV in the last equality.
For the same benchmark with $m_{Z'} = 300$~MeV, the total cross section for upscattering on electrons at UHEs is given by
\begin{equation}
    \sigma_{X}^{e} \simeq 4.2 \times 10^{-35}\text{ cm}^2 \left( \frac{g_X g_e}{10^{-4}}\right)^2.
\end{equation}
One can check that, again, the recoil energy of the target is very small. 
In particular, for the point above, the average electron recoil energy $\langle E_e \rangle \simeq \langle Q^2 \rangle /2 m_e \simeq 40$~GeV.

From the above, we can conclude that as long as the mediator is lighter than several hundred MeV, the upscattering will be mostly invisible inside IceCube's volume.
Only when $m_{Z'} \gtrsim 1$~GeV the DIS and electron scattering regimes start to dominate over QE and coherent upscattering.
From \cref{simFQ}, we observe that as long as  $m_{X’}^2-m_X^2\ll 2E_Xm_N \sim (10~{\rm TeV})^2$, $\sigma_X^{\rm coh}$ and $\sigma_X^{\rm QE}$ are almost independent of the values of $m_{X’}$, $m_X$ and $E_X$. Since we want the subsequent decay of $X’$ to produce the muon pair along with $X$, the splitting should be larger than $2m_\mu$. As we shall see,  $m_{X’}$ of order of $O({\rm GeV})$ is favored both for model building and for the production at the source. The DIS cross section is more sensitive to the mass splitting because the low energy partons corresponding to the parton-$X$ center of mass energy close to $m_{X’}$ dominate DIS. However, for $m_{Z’}\sim 300$ MeV, DIS itself is subdominant. 
At first glance, it may seem counterintuitive that despite the very large center of mass energy, the coherent scattering dominates.  This is of course because of the lightness of the mediator. For electron (or positron) scattering off proton, we have a similar pattern but since both colliding particles and their energy momentum in coherent forward scattering remain almost intact, they are not easily detected.

In deriving \cref{eq:coh-QE-DIS}, we considered a $Z'$ coupled to a baryonic current, where the proton and the neutron couplings are equal and given by $g_B = 3 g_q$. 
Alternatively, $Z’$ can be a dark photon with a kinetic mixing $\varepsilon<10^{-4}$~\cite{NA64:2025ddk}, in which case it couples predominantly to the electromagnetic current.
In that case, $g_e=-g_p = e \varepsilon $ and $g_n = 0$.
The nuclear coupling is then simply $g_N \propto Ze$ and $\sigma^{\rm coh}_X$ obtains an enhancement of $Z$ over scattering off $Z$ electron orbiting a nucleus of $Z$ protons. 
Thus, the scattering over nuclei will dominate for the dark photon model as well. 
For the $B-3L_\mu$ model, the coupling of $Z'$ to the electron at the tree level vanishes and only scattering on nucleus contributes to $X'$ production. 
In principle, if $Z'$ couples only to quarks and the dark particles, $g_q$ can be as large as $10^{-3}$~\cite{Farzan:2016wym}, increasing $\sigma_{\rm tot}$ to $\mathcal{O}(10^{-33})$~cm$^{-2}$.  
However, within the $B-3L_\mu$ model, the oscillation bounds on $\epsilon_{\tau\tau}-\epsilon_{\mu\mu}$ constrains $g_q$ to be smaller than $6\times 10^{-5}$ (or equivalently $g_B < 1.8 \times 10^{-4}$)~\cite{Coloma:2023ixt,AtzoriCorona:2022moj}.

Finally, one can also consider leptophilic models like a gauged $L_\mu-L_e$ model.
The scattering off electrons will then dominate as it is the only allowed interaction at tree-level. 
For concreteness, we restrict ourselves to the scattering off the nuclei throughout the paper, benefiting from the nuclear coherence enhancement and dismissing the $L_\mu-L_e$ model.

\subsection{Metastable particle: $\phi \to \mu^+\mu^-$ \label{sub:Metastable}}

In principle, the same mediator that leads to the $X\to X'$ upscattering  ({\it i.e.,} the dark photon or the gauge boson of $B-3L_\mu$) can also play the role of the metastable particle whose decay produces the dimuon signal. That is, $X'$ would promptly decay to $X+Z'$ via $g_X\sim 0.5$ and then $Z'$ would decay  to the $\mu \bar{\mu}$ pair with a long decay length. 
However, the decay length of the metastable particle should be larger than 100 km which implies $g_B$ or $e\epsilon$ to be smaller than $10^{-6}$. 
This is a value well below the present bound and would suppress the $X$ cross section off nuclei.

While there are many ways to reconcile a sizable cross section with the requirement of a metastable particle, we choose to simply introduce a new mediator, $\phi$, with 
\begin{equation}
    \mathcal{L}_{\rm decay} = y_{\mu} \phi \bar\mu \mu  + y_X \phi (\bar {X'}X + \text{h.c.})
\end{equation}
With the appropriate arrangement of masses for the two mediators and $g_X < y_X$, we can enforce $\Gamma(X'\to Z'+X)/\Gamma(X'\to X \phi)\ll 1$. 
After $X'$ decays to $\phi$, it propagates unimpeded through matter and then decays as $\phi \to \mu^+ \mu^-$ with a typical decay length $\ell_d$ comparable to $L_{\rm KM}$.\footnote{In principle, we could consider lepton flavor violating modes as $\phi \to e^\pm \mu^\mp,\tau^\pm\mu^\mp$. For energies of $O(100~{\rm PeV})$, the propagation range of the $\tau$ would be a few km. 
As a result, we find that the probability of a $\tau$ signal at KM3NeT would be smaller than that of a muon. 
If $\phi$ decays into $e^\pm \mu^\mp$, the produced $e^-$ or $e^+$ will be absorbed before reaching the detector so that only a single muon is observed. Nevertheless, flavor-violating couplings lead to $\mu \to e$ transitions that are strongly constrained 
by various experiments.}
As we shall see in the next subsection, our  solution works best for $L_{\rm KM}\sim \ell_d$ which requires the coupling of $\phi$ to the muon pair to be $y_{\mu} \sim 5 \times 10^{-6}(0.3~{\rm GeV}/m_\phi)$. 
With such a value of coupling between $\phi$ and the $\mu \bar{\mu}$ pair, the contribution to $(g-2)_\mu$ will be far below the present bound \cite{Ge:2021cjz}.   

The coupling of $\phi$ to $\mu \bar{\mu}$ can be obtained by the mixing of $\phi$ with the neutral component of a new Higgs or even the SM Higgs. In the latter case, the mixing should be $y_\mu \langle H\rangle/m_\mu\sim 10^{-3}$. As long as $\phi$ is lighter $2m_K$, the main decay mode will be to dimuon pairs,
\begin{equation}
    B_{\mu\mu}\equiv Br(\phi \to \mu^+\mu^-)\simeq 1.
\end{equation}
In what follows, we focus on $m_\phi > 2m_\mu$, however, we note that if $m_\phi < 2 m_\mu$, the new scalar can still produce dimuon pairs via pair conversion on the Coulomb field of the nucleus, $\phi N\to \mu^+ \mu^- N$, further enhancing the signal yield in rock versus ice.

Finally, we note that if the mass splitting between $X'$ and $X$ is small, $X'$ can play the role of the metastable particle as it can only decay via the three-body process $X' \to X \mu^+\mu^-$ through an off-shell $Z'$. 
Such a scenario would naturally fit into the $B - 3 L_\mu$ model mentioned above and there would be no need for an additional mediator $\phi$.
Our discussion would proceed mostly unchanged, with the replacement $\Gamma_\phi \to \Gamma_{X'}$ and $B_{\mu\mu} \to Br(X' \to X \mu^+\mu^-)$.
For a more detailed discussion of this scenario, see~\cite{Dev:2025czz}.

\subsection{Signal Estimation}
\label{sub:muondetection}

\begin{figure*}[t!]
    \centering
    \includegraphics[width=0.49\textwidth]{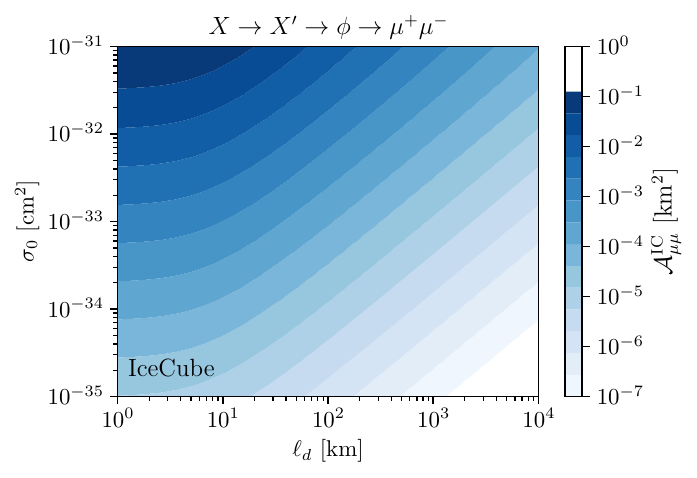}
    \includegraphics[width=0.49\textwidth]{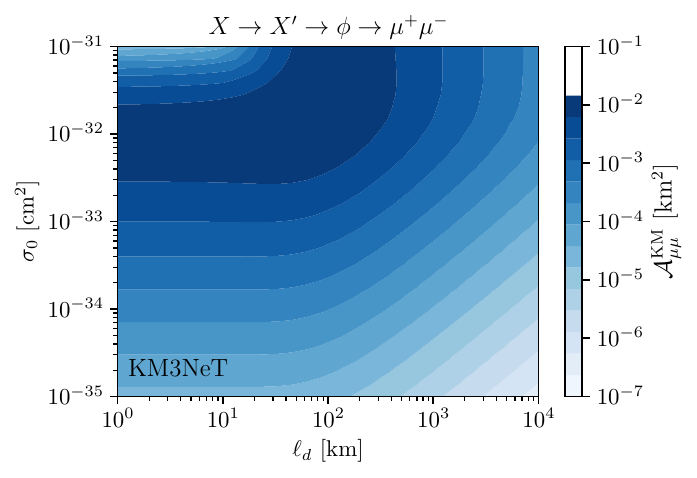}
    \caption{The effective areas of IceCube (left panel) and KM3NeT (right panel) for the $X \to X^\prime \to \phi \to \mu^+ \mu^-$ signal as a function of the upscattering cross section normalization $\sigma_{0}$ (see \cref{eq:sigma0}) and the $X^\prime$ decay length in the lab frame, $\ell_{d}$, assuming a monochromatic source and $\ell_\mu = 10$~km.
    We assume $\theta_{\rm el}^{\rm IC} = 7.8^\circ$ and $\theta_{\rm el}^{\rm KM} = 0.6^\circ$.
    \label{fig:areas}
    }
\end{figure*}

Now we calculate the signal rate for through-going dimuon tracks at KM3NeT.
At such ultra high energies all particles are extremely boosted so we are justified in simplifying the signal estimation as a one-dimensional problem.
That is, once the dark particle source location is fixed, the line of sight of the propagating $X$ particle is fixed and the probability of detection is approximately a function of a single spatial variable.
We can therefore safely assume that only the $X$ particles for which the continuation of their paths cross the detector may lead to a signal. 

Denoting the time integrated flux with $\phi_X$, the numbers of such $X$ particles that intersect the KM3NeT and IceCube volumes are respectively equal to $\phi_X {\rm A_{\rm KM}}$ and $\phi_X {\rm A_{\rm IC}}$. 
The number of dimuon events at a given detector ${\rm D= KM, IC}$ can then be estimated as
\begin{equation}
    N_{\mu\mu}^{\rm D} = \phi_X \mathcal{A}_{\mu \mu}^{\rm D},
\end{equation}
where $\mathcal{A}^{\rm D}_{\mu \mu}$ is the effective area of the detector to the $X\to X'\to \phi \to \mu^+\mu^-$ signal, such that the required flux to explain the \kmevent event at KM3NeT is just 
\begin{equation}\label{phiINVA}
    \phi_X = \left(\mathcal{A}_{\mu \mu}^{\rm KM}\right)^{-1}.
\end{equation}

Since virtually every single isolated high-energy muon (or collimated dimuon pair) that crosses KM3NeT or IceCube is detected at these energies, the effective areas relevant for our scenario can be crudely approximated as the probability of detecting the dimuon pair times the detector transverse area ${\rm A_D}$.
We treat the two detectors as cylinders of diameter $d$ and height $h$, we take $(d,h) = (1.1~\text{km},1~\text{km})$ for IceCube and $(d,h) = (0.52~\text{km},0.7~\text{km})$ for ARCA-21.
The corresponding transverse areas are then ${\rm A_{IC}} \simeq 1.1$~km$^2$ and ${\rm A_{\rm KM}}\simeq 0.36$~km$^2$.
Track quality and track length selection criteria decrease these values somewhat, but do not invalidate our main argument.

The probability of detecting the dimuon signal from the upscattering chain $X \to X^\prime$, $X'\to X\phi$, and then $\phi\to \mu^+\mu^-$ is the product of the probability of scattering, decay(s), and muon survival along the line of travel of $X$.
Since $X' \to X\phi$ is prompt, we take its decay probability to be unity, and find
\begin{align}\label{eq:Amumu}
    \mathcal{A}_{\mu\mu}^{\rm D} = {\rm A}_{\rm D} &\int_0^{L_{\rm D}}  \frac{\dd y}{\ell_d}  \int_0^y \frac{\dd x}{\ell_s(x)}  \, \exp\left(- \int_{0}^{x}\frac{\dd u}{\ell_s(u)}\right)
    \\\nonumber
    &\times 
    \exp\left( - \frac{y-x}{\ell_d}\right) 
    \exp\left(-\frac{L_D - y}{\ell_\mu}\right) B_{\mu\mu}.
\end{align}
Here, $\ell_s(x) = \left[n(x) \sigma_X(E_X)\right]^{-1}$ is the scattering length of $X \to X'$ in the material with target number density $n(x)$.
The decay length of a metastable particle of energy $E$ produced in $X'$ decays is $\ell_d  = \Gamma^{-1}(E)$ in the lab frame and $\ell_\mu = \ell_\mu(y)$ is the muon propagation range in the material.
Finally, the branching ratio of the metastable particles into muons is $B_{\mu\mu}$,  which for $\phi \to \mu^+\mu^-$ is simply equal to 1.
The distance between the surface and the beginning of the detector region along the direction of travel of $X$ is denoted by $L_{\rm D}$.
As an approximation, we take every $\phi$ decay that occurs at a distance of $\ell_\mu$ from the detector to be considered a through-going track signal.

The expression in \cref{eq:Amumu} neglects any angular deviation in the propagation of the particles.
This is a good approximation because as emphasized before the upscaterring of $X$ is strongly peaked in the forward direction. Moreover,    the angular separation between daughter particles in the decays $X'\to X \phi$ and $\phi \to \mu \bar{\mu}$ is suppressed by the ratio of the mother mass to its energy which is smaller than $ 10^{-8}$ and therefore completely negligible.
The energy loss by  muon daughter particles in the medium may cause a larger  angular deviation.
However, this effect is far smaller than the typical spatial resolution of both IceCube and ARCA-21.
From \cref{eq:coh-QE-DIS}, we conclude that the probability of rescattering, such as $X\to X' \to X \to X'$, is negligible.
 
Let us first consider the case of constant density. As discussed earlier, the mean free path of the scattering is much larger than the Earth diameter:
 $$ \frac{L_{\rm D}}{\ell_s}=n \sigma_X L_{\rm D}< n\sigma (2R_\oplus)\ll 1.$$
The dimuon effective area can be  then written as 
 \begin{equation} \label{fXwithA}
    \mathcal{A}_{\mu\mu}^{\rm D} = {\rm A}_{\rm D}(n \sigma_X L_{\rm D}) f_X,
\end{equation} where, as defined before, $f_X$ is the probability that  $X'$ from $X$ scattering leads to a signal. Taking $\ell_d,L_{\rm D}\gg \ell_\mu$, from \cref{eq:Amumu}, we obtain 
\begin{equation} f_X=\frac{\ell_\mu}{L_{\rm D}}(1-\exp (- L_{\rm D}/\ell_d)). \label{fXX}\end{equation}
For $\ell_d\sim L_{\rm D}=L_{\rm KM}=140$~km and $\ell_\mu \sim$ few km, $f_X$ can be as large as $0.01$.


In this limit, given that most of the baseline at KM3NeT is made of rock, the ratio between IceCube and KM3NeT events is approximately,
\begin{align}\label{eq:ratioICEtoKM}
    R_{\rm IC/KM} &\equiv \frac{\mathcal{A}_{\mu\mu}^{\rm IC}}{\mathcal{A}_{\mu\mu}^{\rm KM}} 
    \\\nonumber
    &\simeq \left(\frac{\sigma_X^{\rm H_2O}}{{\sigma_{X}^{\rm SiO_2}}}\frac{m_{\rm SiO_2}}{m_{\rm H_2O}}\right) \frac{\rho_{\rm ice}}{\rho_{\rm rock}} \frac{\rm A_{IC}}{\rm A_{\rm KM}}\frac{L_{\rm IC}}{L_{\rm KM}} \simeq { 0.07},
\end{align}
where $L_{\rm IC}/L_{\rm KM} \simeq 0.1$ and we assumed the muon range to be the same.
The ratio of the upscattering cross sections in rock and ice was taken to be approximately $\frac{\sigma_X^{\rm H_2O}}{{\sigma_{X}^{\rm SiO_2}}}\frac{m_{\rm SiO_2}}{m_{\rm H_2O}} \simeq \frac{2}{3}$, which is approximately true for both the case of a dark photon and a baryonic current like $B - 3L_\mu$.
\Cref{eq:ratioICEtoKM} demonstrates that with a transient source within our model, IceCube would observe null result. This was the aim of the present paper.

The energy dependence in each factor is also implicit, but to a good approximation, 
\begin{equation}
    E_X \simeq E_{X'} \simeq 2 E_\phi \simeq 4 E_{\mu^\pm}.
\end{equation} 
The source of $X$ most likely emits a continuous energy spectrum, $F_X(E_X)$, rapidly decreasing at energies above 200 PeV. As discussed before, the energy dependence of the $X$ upscattering cross section and therefore that of $\ell_s$ is mild. However, due to the relativistic boost factor of the decay rate, $\ell_d$ is proportional to $E_X$. The muon range decreases with decreasing the energy. Taking $\ell_d\sim L_{\rm D}=L_{\rm KM}$ at 100 PeV, for $E_\phi<10$~PeV, $f_X=\ell_\mu/L_{\rm D}\ll 0.01$. Thus, the number of events would be suppressed by $\mathcal{A}_{\mu\mu}^{\rm KM}\propto f_X$ at lower energies. To explain the null results of KM3NeT at lower energies, the energy spectrum of $X$, $F_X(E_X)$, should be hard such that for $100~{\rm TeV}<E_X<200~{\rm PeV}$, $F_X(E_X)/\ell_\mu(E_X/4)$ should be an increasing function of $E_X$.

\Cref{fig:areas} shows the effective areas $A_{\mu \mu}^{\rm D}$ for KM3NeT and IceCube in the nominal direction of \kmevent as a function of the decay length of $\phi$, $\ell_d$, and of a proxy for the upscattering cross section defined by, 
\begin{equation}\label{eq:sigma0}
    \sigma^{\rm coh}_X(A) = \sigma_0 A^2
\end{equation}
where $\sigma^{\rm coh}_X(A)$ is the total upscattering cross section on a nucleus of mass number $A$.
The above assumes that the cross section is dominantly in the coherent regime, which as discussed in \cref{sub:scattering} is a good approximation for our models.
For dark photon models, the scaling is instead proportional to $Z^2$, although the impact on the ratio $R_{\rm IC/KM}$ is very small.
In drawing these effective areas, we assume a constant $\ell_\mu = 10$~km, such that the dimuon pair should maintain the bulk of its $E_{\mu^+} + E_{\mu^-} \simeq E_X/2$ energy.
In the case of IceCube, the column depth of ice is never large enough to become opaque to $X$ particles within the cross section range shown.
However, at KM3NeT, the effective area peaks at about $\sigma_{X}\simeq 10^{-31}$~cm$^2$ and $\ell_d\simeq 100$~km, as at that point most $X$ particles interact before reaching the detector and decay within $L_{\rm KM}$.

\begin{figure*}[t]
    \includegraphics[width=0.49\textwidth]{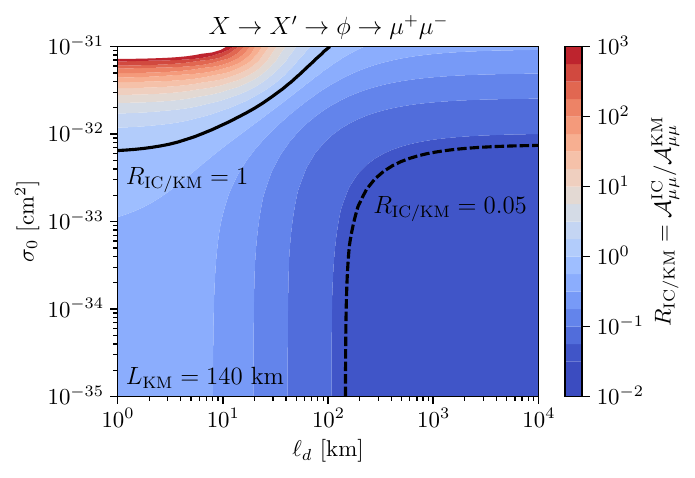}
    \includegraphics[width=0.49\textwidth]{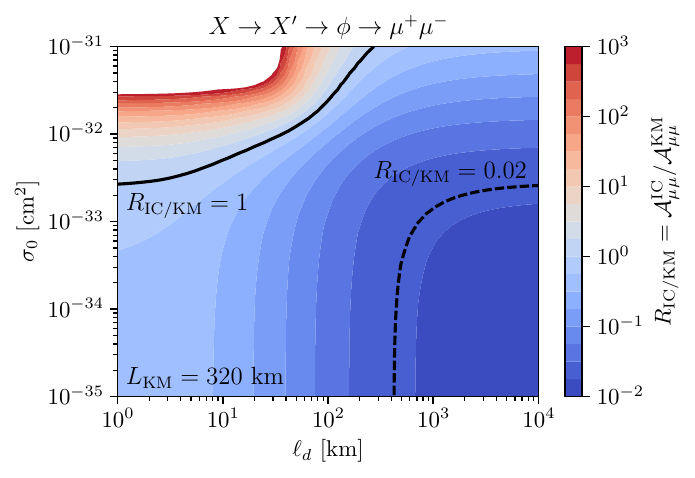}
    \caption{
    The ratio between the IceCube and KM3NeT number of events for $\theta_{\rm el}^{\rm KM} =0.6^\circ$ (left) and $\theta_{\rm el}^{\rm KM} =-0.9^\circ$ (right) as a function of the total upscattering cross section normalization $\sigma_{0}$ (see \cref{eq:sigma0}) and the $X^\prime$ decay length in the lab frame, $\ell_{d}$, assuming a monochromatic source.
    The solid lines correspond to a ratio of $1$ and the dashed line to a value of $0.05$ on the left and $0.02$ on the right, respectively.
    \label{fig:ratio}
    }
\end{figure*}

\Cref{fig:ratio} shows the ratio between the IceCube and KM3NeT effective areas of \cref{eq:ratioICEtoKM} for a monochromatic source assumption for two different KM3NeT elevation angles and fixing $\ell_\mu = 10$~km.
This serves as a proxy for the ratio of the number of events for an instantaneous burst.
Since IceCube's angle is less sensitive to the angular uncertainty, we maintain it fixed at $\theta_{\rm el}^{\rm IC} = 7.8^\circ$.
The result is as expected: the ratio in the number of events is maximized in the regime of \emph{long lifetimes}, $\ell_d \gtrsim 10^2$~km, and \emph{small cross sections}, $\sigma_0 \lesssim 10^{-32}$~cm$^2$.
In that regime, the ratio is roughly constant, as expected from \cref{eq:ratioICEtoKM}. This  implies that our scenario for reconciling the  null signal at IceCube with the  KM3-230213A  event is robust against the uncertainties in the determination of the elevation of the source.

\subsection{Duration and  magnitude of the $X$ flux\label{sub:duration}}

So far, we have implicitly assumed that the emission of $X$ by the source is instantaneous, or at least shorter than a few hours, such that the path length traversed by $X$ on its way to the detectors remains constant.
For longer dark bursts, as the Earth spins around, the length $L_{\rm KM}$ at KM3NeT, located at a latitude of $37^\circ$ N, varies significantly.
IceCube, however, is located at the South Pole, so $L_{\rm IC}$ remains relatively constant throughout the day. 
With cross sections of few$\times 10^{-34}$ cm$^2$ (see \cref{eq:coh-QE-DIS}), the mean free path in Earth is larger than its radius which means the probability of the $X$ scattering before KM3NeT grows linearly with $L_{\rm KM}$. 
However, combining Eqs. (\ref{fXwithA},\ref{fXX}), we conclude that for $\ell_d\sim 100$~km, as $L_{\rm KM}$ grows from 140 km to $2 R_\oplus$, $\mathcal{A}_{\mu\mu}^{\rm KM}$ increases only 40 \%.  This means that with $\ell_d\sim 100$ km and  a mild time variation of the incoming $X$ flux, the predicted rate of events at KM3NeT will not considerably change throughout the time that the source is below the  horizon of KM3NeT. 
 For larger decay length, $\ell_d\sim R_\oplus$, $\mathcal{A}_{\mu\mu}^{\rm KM}$ increases by a factor $O(10)$ or larger as $L_{\rm KM}$ grows from 140~km to $2R_\oplus$. This means with $\ell_d\gsim 1000$ km, we expect more events at KM3NeT as the source drops well below the horizon which is in contradiction with the observation. 
 
In conclusion, as long as the duration of the source is smaller than the data taking time of KM3NeT, it does not matter whether the $X$ particles come as a burst or a longer duration transient flux. 
However, the fact that the event is registered when $L_{\rm KM}\sim$140 km favors $\ell_d \lesssim \mathcal{O}(400)$~km and strongly disfavors $\ell_d>\mathcal{O}(10^{3})$~km.
Regardless of the time shape of the flux, as long as it is transient with duration less than KM3NeT data taking duration, the predicted number of events in IceCube remains much smaller than that at KM3NeT and therefore consistent with observation.

These considerations along with the relation \cref{phiINVA} point towards a total energy and time integrated flux of
\begin{equation} 
\label{phi-X}
{\phi_X \simeq 2} \times 10^{-4} \text{ cm}^{-2} \left(\frac{10^{-4}}{g_Xg_B}\right)^2 \left( \frac{0.01}{\ell_\mu/\ell_d} \right).
\end{equation}

\section{Speculations on the  Source}
\label{sec:sources}

In the literature, dark matter decay has been suggested as a source for KM3-230213A \cite{Borah:2025igh,Kohri:2025bsn,Khan:2025gxs,Murase:2025uwv,Barman:2025hoz}. 
However, such a source is not transient and cannot reconcile the null results from IceCube with KM3-230213A. 
A transient source such as a choked gamma ray burst (GRB)~\cite{Meszaros:2001ms} or a transient flare in active galactic nuclei (AGN) will be a more suitable candidate for the source of the KM3-230213A event. 
The $pp$ or $p \gamma$ collisions at the source can produce $X$ and $X'$ via the $Z'$ mediator. 
However, such collisions also produce $\gamma$ ray and neutrino fluxes with much higher luminosity. 
Refs.~\cite{Fang:2025nzg,Crnogorcevic:2025vou} show that the intergalactic magnetic field can eliminate the gamma ray flux, in accordance with the null signal by Fermi-LAT. 
Working out the details of the dynamics of the source is beyond the scope of the present paper but in the following, we suggest a mechanism through which $\nu_L$ can be converted to the messenger that induces the KM3-230213A event. 

Let us introduce the following dipole interaction term
\begin{equation} 
\mathcal{L}_\nu \supset \mu_\nu \bar{\nu}_R\sigma^{\alpha \beta} \nu_L F_{\alpha \beta} + \text{ h.c.}
\end{equation}
Upper bounds on $\mu_\nu$ come from signatures of solar neutrinos and astrophysics, but values of $\mu_\nu \simeq 10^{-12}\mu_B=10^{-12} e/(2m_e)$ are currently allowed~\cite{XENON:2022ltv} (see also~\cite{Brdar:2020quo}). 
The active sterile oscillation probability in presence of a coherent magnetic field  $B_\perp$ perpendicular to the momentum of the neutrino can be written as~\cite{Kopp:2022cug}
\begin{equation} \label{LRosci}
    P(\nu_L \to \nu_R) =\sin^2 \frac{\mu_\nu B_\perp L}{2}.
\end{equation}
This relation holds valid for (quasi-)degenerate $\nu_L$ and $\nu_R$.
As shown in \cite{Marti-Vidal:2015fdk}, inside the jets of AGN, the magnetic field can be as strong as 10 Gauss with a coherence length of 0.01 pc which implies that for $\mu_\nu>10^{-14} \mu_B$,  the conversion of $\nu_L$ to $\nu_R$ can be significant.  The product of magnetic field times  the coherence length inside a GRB source can be even larger \cite{Piran:2005qu}, yielding a significant conversion rate in a GRB source, too. However,  $\mu B_\perp L/2 =n\pi/2$ ($n\in Z$) and therefore $P(\nu_L \to \nu_R)=1$ requires an unlikely fine tuning. 

If in front of the neutrino and photon flux of energy of 100 PeV, a clump of mass with $\int \rho \,dL\sim 200$ t/cm$^2$ (less than the Earth column density when traversing the Earth diameter) exists, both fluxes can completely eliminated, conforming with the null signal from IceCube as well as from Fermi-LAT. 
Considering the enormity of the source, the existence of such a clump of matter is not unreasonable. While $\nu_R$ can pass through such a clump unattenuated, the $\nu_L$ clump cannot.
However, due to the intergalactic or inter-stellar magnetic field, the $\nu_R$ flux can partially convert back to $\nu_L$, unless  the average perpendicular magnetic field along the distance of $\sim$ Gpc between the source and the milky way galaxy (inside the milky way) is much smaller than $10^{-10}$ Gauss (than $10^{-5}$ Gauss). 
As  will be shown in the next section, the $\nu_R$  mixing with $X$ (as demonstrated  in \cref{eq:matrix}) yields a coupling of form in \cref{eq:generic_couplings} which leads to the KM3-230213A signal by producing $X’$ in the scattering of $X$ off the nuclei in the Earth.

Let us now evaluate  the energy released in the form of messengers ($X$ or $\nu_R$) that lead to the single KM3-230213A event. Taking the solid angle of the flux $\Omega=2\pi(1-\cos \theta_o)$ with $\theta_o=23^\circ$ (corresponding to the average opening angle of AGN jets \cite{Pushkarev:2017fbk})  and the luminosity distance of the source $D=1$ Gpc, we can estimate the energy released as
\begin{align}
&E_{\rm source}  = \Omega D^2 \phi_X E_X = 
\\ \nonumber
&2.5 \times 10^{56} ~{\rm erg} ~\left(\frac{E_X}{200~{\rm PeV}}\right) \left( \frac{\phi_X}{2\times 10^{-4} {\rm cm}^2}\right),
\end{align}
where we have used the estimate for the flux in Eq. (\ref{phi-X}).
This energy release is well below the energy of most energetic registered AGN  \cite{Gitti:2007pg}, where $E_{\rm source} \sim 6 \times 10^{61}$ erg. 
For RGBs, the opening of the jet can be even smaller.
Inserting the opening angle computed in \cite{Goldstein:2015fib} for RGB, we obtain $10^{54}$ erg. 
The energy of GRB221009A was estimated to be higher than this value~\cite{Frederiks:2023bxg}.
It seems both GRBs and AGNs can  have enough energy and can moreover have sufficiently large magnetic fields to convert $\nu_L$ to $\nu_R$ (or to $X$) to explain the KM3NeT observation in our scenario.
The material around the GRB source can ``choke" the produced $\gamma$-ray and the active neutrino flux, explaining the absence of accompanying channels.

\section{Model Details}
\label{sec:models}

Let us first discuss how $Z'$ can obtain a coupling of form $g_X\bar{X} \gamma^\mu X' Z'_\mu$. 
Following Ref.~\cite{Farzan:2019xor}, we define
$$ \psi_1=\frac{X+X'}{\sqrt{2}} \ \ \ {\rm and} \ \ \psi_{2}=\frac{X-X'}{\sqrt{2}}.$$
Remember that $Z'$ is the gauge boson of a new $U(1)$ symmetry. Assigning opposite charges to $\psi_1$ and $\psi_2$, the $Z'$ couples to the dark current
\begin{align} \label{eq:gX}
(\bar{\psi}_1 \gamma^\mu \psi_1 -
\bar{\psi}_2 \gamma^\mu \psi_2) = (\bar{X'}\gamma^\mu X+
\bar{X}\gamma^\mu X')
\end{align}
We therefore obtain off-diagonal  coupling for $X$ and $X'$ to $Z'$. This is not however sufficient for our scenario as we should ensure $X$ and $X'$ are mass eigenstates with a mass splitting. Notice that, regardless of  whether $\psi_1$ and $\psi_2$ are Dirac or Weyl spinors, the contributions from $\psi_1$ and $\psi_2$ to the gauge anomaly cancel out thanks to their opposite $U(1)$ charges.
In the following, we discuss both Majorana and Dirac mass mechanisms for $X$ and $X'.$
\begin{itemize}
\item The $U(1)$ symmetry allows mass terms of the following form 
$$ M(\psi_1^T c\psi_2+\psi_2^Tc\psi_1)= M(X^TcX-X'^Tc X')$$
which gives equal masses to $X$ and $X'$. The challenging part is to induce mass splitting between $X$ and $X'$ which requires spontaneous breaking of the $U(1)$ gauge symmetry.
One economic solution is to introduce a coupling of the following form with a new scalar with twice the $U(1)$ charge 
of $\psi_2$:
\begin{align}
     \mathcal{L} & \supset -\lambda (\Phi \psi_1^Tc\psi_1+\Phi^* \psi_2^Tc\psi_2)
     \\
     & = -\lambda\Big[\Phi_R(X^TcX+X'^TcX') 
     \\\nonumber & 
     \qquad \qquad + i \Phi_I (X'^TcX+X^TcX') \Big],
\end{align}
where $\Phi=(\Phi_R+i \Phi_I)/\sqrt{2}.$
Similarly to Ref. \cite{Farzan:2019xor}, the equality of the couplings to $\psi_1$ and $\psi_2$ can be justified by a symmetry under the following transformation
\begin{equation} \label{1To2}
\psi_1 \leftrightarrow \psi_2 \ , \ Z' \leftrightarrow -Z', \ {\rm and} \ \ \ \Phi  \leftrightarrow \Phi^*. 
\end{equation}
As for the SM Higgs, we can go to the unitary gauge where $\langle \Phi \rangle$ and $\Phi$ are real, giving rise to  $$m_{X}=M-\lambda \langle \Phi\rangle \ {\rm and} \ m_{X'}=M+\lambda \langle \Phi\rangle.$$
Thus, $m_{X'}^2-m_X^2=4\lambda M\langle \Phi\rangle.$
The vacuum expectation value of $\Phi $ 
also gives rise to the $m_{Z'}$ mass:
$$ m_{Z'}=2g_X \langle \Phi\rangle.$$
    We  therefore find
\begin{equation}
    g_X= \lambda \times \left(\frac{m_{Z'}}{m_{X'}-m_X}\right).
\end{equation}
\item For Dirac $\psi_1$ and $\psi_2$, we can introduce terms $\bar{\psi}_1\psi_1 $ and $\bar{\psi}_2 \psi_2$ which respects the new $U(1)$.  Let us make a slight change of notation and redefine $\psi_1=(\hat{X}+X')/\sqrt{2}$ and $\psi_2 =(\hat{X}-X')/\sqrt{2}$. Moreover let us denote the gauge coupling with $g_\psi$: $g_\psi(\bar{\psi}_1\gamma^\mu \psi_1-\bar{\psi}_2 \gamma^\mu \psi_2) Z'_\mu.$ Imposing a $Z_2$ symmetry under which $\psi_1\leftrightarrow \psi_2$ and $Z' \leftrightarrow -Z'$, we obtain equal masses for $\hat{X} $ and $X'$: $$ M(\bar{\psi}_1\psi_1+\bar{\psi}_2\psi_2)=
M(\bar{\hat{X}}\hat{X}+\bar{X'}X').$$
Then, $\hat{X}$ and $X'$ would be degenerate. Let us introduce a right-handed neutrinos, $\nu_R$ which is a gauge group singlet. Moreover, we  introduce a new scalar $\Phi$ with  a charge equal to that of $\psi_1$. Under the $Z_2$ symmetry, $\Phi \leftrightarrow \Phi^*$. We can then write the following Yukawa couplings
$$  \lambda_R \bar{\nu}_R(\Phi^* \psi_1+\Phi \psi_2)+{\rm H}.c.$$
Similarly to the SM and   to the case discussed in the first item, by going to the unitary gauge, we can make $\Phi$ real. Then, ${\nu}_R$ can mix with $\hat{X}$ (but not with $X'$):
$$ \sqrt{2} \lambda_R \langle \Phi \rangle \bar{\nu}_R \hat{X}_L +{\rm H}.c.$$  After the electroweak symmetry breaking, $\nu_R$ can form a Dirac mass term with the active left-handed neutrinos. We therefore obtain
\begin{align} \label{eq:matrix}
[\bar{\nu}_L \ \bar{\hat{X}}_L] \left[\begin{matrix} {m}_\nu & 0 \cr \sqrt{2}\lambda_R\langle \Phi\rangle & M\end{matrix} \right]\left[\begin{matrix}  \nu_R \cr \hat{X}_R \end{matrix}\right]. 
\end{align}
Taking ${m}_\nu\ll \sqrt{2} \langle \Phi\rangle \lambda_R , M$, we obtain a mass eigensystem as follows:
\begin{align}
X_R &\simeq \cos \alpha~\nu_R + \sin\alpha \hat{X}_R, 
\\\nonumber
\tilde{\nu}_L &\simeq \nu_L + \beta X_L, 
\end{align}
with mass
\begin{align}
\tilde{m}_\nu \simeq m_\nu \left(1 - (\frac{\sqrt{2} \lambda_R \langle \Phi \rangle }{M})^2 \right), \notag 
\end{align}
and
\begin{align}
\tilde{X}_R &\simeq \cos \alpha~\hat{X}_R -\sin \alpha \nu_R,
\\\nonumber
X_L &\simeq \hat{X}_L - \beta \nu_L, 
\end{align}
with mass $M$, in which 
\begin{align}
\sin \alpha &= \frac{\sqrt{2}\lambda_R \langle \Phi\rangle }{\sqrt{M^2+2\lambda_R^2 \langle \Phi\rangle^2 }}, 
\\\nonumber
\beta &= m_\nu \frac{ \sqrt{2}\lambda_R \langle \Phi\rangle}{M^2}\ll  1.
\end{align}
Notice that in the limit $m_\nu \to 0$, $\tilde{\nu}_L$ and $X_R$ will be massless and $\tilde{\nu}_L$ does not mix with $X_L$. 
Indeed, $\nu_L$ is irrelevant for our discussion and the mass mechanism for active neutrinos may come from another mechanism altogether. 
However, the mixing between $\nu_R$ and $X_R$ has a very interesting implication for our model. 
Through this mixing, we obtain 
\begin{equation} \label{gR} 
g_X(\bar{X}'_R \gamma^\mu {X}_R+\bar{{X}}_R \gamma^\mu X'_R)Z'_\mu
\end{equation}
with
\begin{align} 
g_X&= g_\psi\sin\alpha 
\simeq 0.42 \lambda_R \times \left(\frac{m_{Z'}}{0.3~{\rm GeV}}\right)\left(\frac{1~{\rm GeV}}{M}\right),
\end{align}
where we identify $g_\psi \langle \Phi\rangle = m_{Z'}$.
\end{itemize}
In both of the above scenarios, we impose a discrete symmetry under which $Z'$ is odd.  
Of course, the couplings of $Z'$ to the quarks and leptons break this discrete symmetry. 
Thus, this approximate symmetry can be considered a rationale for $g_q\ll g_X$.
 
The metastable particle introduced in our main scenario is a scalar with couplings of form $\phi \bar{\mu} \mu$ and $\phi \bar{X}X'$. 
As a $U(1)$ singlet, a $\phi$ coupling of the form $\phi \bar{X}X'$ can be obtained from the following $U(1)$ singlet combination which is also invariant under transformation $\psi_1\leftrightarrow \psi_2$ (see \cref{1To2}) and $\phi \leftrightarrow -\phi$,
\begin{equation}
y_X \phi (\bar{\psi}_1\psi_1-\bar{\psi}_2\psi_2)=y_X\phi(\bar{X}X'+\bar{X}'X).
\end{equation}

As alluded to before, the $\phi$ coupling to muons can be obtained by mixing $\phi$ with the SM Higgs. 
Such a mixing can come from a simple trilinear coupling $\phi |H|^2$ which respects the gauge group but softly breaks the symmetry under $\phi \leftrightarrow -\phi$.


\subsection{Cosmological implications}
\label{sec:cosmo}

The new particles introduced in this model can all be produced in the early Universe. Let us divide them to three groups: (1) $X$ and $X’$ with  masses of $\sim 1$ GeV or larger. (2) $Z’$, $\Phi$ or $\phi$ with masses of few 100 MeV; (3) $\nu_R$ with mass equal to the active neutrino mass which may contribute to the  extra relativistic degrees of freedom.  
$X’$ can decay fast to $X$. With  $\langle \sigma(X\bar{X} \to Z’Z’)v\rangle \sim (g_X^4/16\pi)(m_X^2/m_{X'}^4)\sim 5 \times 10^{-31}~{\rm cm}^2 (g_X/0.5)^4 (1~{\rm GeV}/m_{X'})^4 (m_X/1~{\rm GeV})^2 $, the contribution of $X$ to dark matter [({\it i.e.,} $\Omega_X/\Omega_{DM}\sim (1~pb)/\langle \sigma(X\bar{X} \to Z’Z’)v\rangle$] will be less than $10^{-5}$. $Z’$ and $\phi$ can decay to the SM leptons before neutrino decoupling era as $Z’\to \nu_\mu\bar{\nu}_\mu, \mu\bar{\mu}$, $\phi \to \mu \bar{\mu}$. Similarly, $\Phi$ can promptly decay to a $Z’$ pair. 

$\nu_R$ can be produced  in the early universe via the interaction with $Z’$ and $X’$ (but not via the dipole moment with $\mu_\nu<10^{-12}\mu_B$) at temperatures above the $X’$ mass (that is above $\sim 1$ GeV). However, its contribution to the extra relativistic degrees of freedom will be suppressed because of the entropy pump by decoupling degrees of freedom in the lower temperatures. For each $\nu_R$, the contribution to $N_{eff}$ will be $N_s^a/N_s^b$ where $N_s^a$ and $N_s^b$ are the entropic degrees of freedom for $1~{\rm MeV}<T<100~{\rm MeV}$ and $100~{\rm MeV}<T<$GeV, respectively. While $N_s^a$ accounts for only the electron, three active neutrinos and the photon, $N_s^b$ also includes $\mu$, three colors of light quarks ($u,d,s$) and eight gluons.  Thus, $\Delta N_{eff}=n N_s^a/N_s^b=0.17 n$ where $n$ is the number of $\nu_R$ in the model.
The present uncertainties on $N_{eff}$ \cite{Planck:2018vyg} therefore allows for one species of $\nu_R$ ($n=1$) but not more. We can in principle have dipole interaction of all three left-handed neutrinos with a single $\nu_R$ and therefore $\nu_{iL}\to \nu_R$ conversion in the magnetic field. However, since $\nu_{iL}$  have mass splitting among themselves, this would require mass splitting between (at least one of) $\nu_L$ and $\nu_R$ of order of $\Delta m_{atm}^2$. As long as $\Delta m_{atm}^2/4 E_\nu \ll \mu_\nu B_\perp$, we can still use Eq. (\ref{LRosci}) but for smaller values of $\mu_\nu B_\perp$, the conversion probability is suppressed by $4 E_\nu^2\mu_\nu^2B_\perp^2/\Delta m_{atm}^2$. Within an AGN with $B_\perp\sim 10$ Gauss, we can still have sizable conversion rate for $\mu_\nu\sim 10^{-12} \mu_B$. In chocked GRB source, the magnetic field is very high \cite{Piran:2005qu} so the condition 
$\Delta m_{atm}^2/4E_\nu\ll \mu_\nu B_\perp$ can be fulfilled for the whole range of $\mu_\nu$ that $\mu_\nu B_\perp L \sim 1.$

\subsection{Discovery  by terrestrial experiments}
In our model, $Z'$ couples to the quarks with a coupling close to the present LHCb bound \cite{LHCb:2019vmc}. Not surprisingly, further data taking by the LHCb can probe regions of parameter space that yield sizeable $X$ scattering cross section off nuclei and provides a plausible explanation for the KM3NeT event. If $Z'$ is the dark photon, it can decay as $Z'\to e^-e^+$. As shown in \cite{Craik:2022riw}, for $m_{Z'}<400$~MeV, the LHCb search for $Z'\to e^-e^+$ with the total run 3 data can probe kinetic mixings down to $7\times 10^{-6}$ which is more than ten times smaller than the present bound.

LHCb can also test the $B-3L_\mu$ scenario for which $Z' \to \mu^-\mu^+,\nu_\mu\bar{\nu}_\mu$. The recently proposed SHIFT@LHC experiment \cite{Niedziela:2024khw}
is also suitable for this purpose. However, as mentioned earlier searching for neutral current Non-Standard Interactions (NSI) at the neutrino oscillation experiments as well as at the Coherent Elastic $\nu$ Nucleus experiment yield the best bound on the coupling for the $B-3L_\mu$ model. In particular, studying high energy atmospheric neutrinos at the neutrino telescopes such as KM3NeT/ORCA \cite{KM3NeT:2024pte} or ICECUBE/DeepCore \cite{IceCubeCollaboration:2021euf} has a great potential to probe small values of $\epsilon_{\mu\mu}-\epsilon_{\tau\tau}$ and therefore the coupling of $B-3L_\mu$. If these experiments further tighten the bound on the coupling, we can consider $B-3(L_\mu+L_\tau)/2$ or $B-L$ which yield zero $\epsilon_{\mu\mu}-\epsilon_{\tau \tau}$.

Light gauge bosons coupled to the SM fermions  appear in a myriad of models. The discovery of such a particle would not necessarily validate our model. The smoking gun for our model is a pair of new relatively light neutral pair of particles with mass below GeV, $X$ and $X'$, that couple to $Z'$. The $\bar{X}X'$ or $X'\bar{X}$ pair can be produced at the $pp$ collisions with a cross section of 
$$\sigma_{pp}(pp\to \bar{X}X')=\sigma(pp\to \bar{X}'X)$$ $$\sim \frac{(g_Xg_q)^2}{4\pi}\frac{0.25^2}{s}\frac{\log(s/m_{XX'}^2)-1}{(m_{XX'}^2/s)^{1.1}},$$
in which $s$ is the square of the center of mass energy of the colliding protons and $m_{XX'}$ is the invariant mass of the final $X$ and $X'$ pair.  We have used the results of \cite{Gayler:2002ve} for the parton distribution function.
For $g_Xg_q\sim 3\times 10^{-5}$ and $m_{XX'}\sim$GeV ($m_{XX'}\sim$10 GeV), $\sigma_{pp}\sim 200$ fb (2 fb). The produced $X'$ (or $\bar{X}'$) will promptly decay to $X$ (to $\bar{X}$) and $\phi$. The final $X\bar{X}$ pair will appear as missing energy but $\phi$ leads to a displaced vertex of a $\mu\bar{\mu}$ pair. Remember that the decay length of $\phi$ in our model is $\ell_d\sim 0.02~{\rm cm} E_\phi/m_\phi$ (such that $E_\phi \sim 100$ PeV, $\ell_d\sim 200$~km) so the displacement of the $\mu\bar{\mu}$ vertex at colliders with $E_\phi\sim m_{XX'}\sim$few GeV will be of order of 0.1~cm. This displacement is too short to be resolved in the beam dump experiments but, being of order of the $B$ meson decay length used for the $b$-tagging, can in principle be resolved at  the LHC detectors such as CMS. With 300 fb$^{-1}$ (3000 fb$^{-1}$) of data during run 3 (during HL-LHC), we shall have $6 \times 10^4$ events ($6 \times 10^5$ events) with displacement of 0.1~cm and 600 events (6000 events) with displacement of $\sim 1$~cm. The invariant mass of the $\mu\bar{\mu}$ pair will peak at $m_\phi$ which provides another handle to reduce the background. Evaluating the quantified discovery is beyond the scope of the present paper.

\section{Conclusions}
\label{sec:conclusions}

We presented a new physics interpretation of the ultra-high-energy (UHE) muon at the ARCA detector of KM3NeT based on transient astrophysical sources of dark particles.
In our scenario, the muon signal arises from a combination of scattering and decay processes as dark particles traverse the interior of the Earth, making its detection highly sensitive to the column depth seen by the incoming flux. 
At the time of the \kmevent event, the new particles traverse a column depth of $\mathcal{O}(7 - 70)$~t/cm$^2$ before reaching the detector, substantially enhancing the likelihood of through-going muon track signals. 
This contrasts with the $\mathcal{O}(1-2)$~t/cm$^2$ column depth at IceCube, explaining why the same source was not observed there.

With the proposed dark particle signal, the tension between KM3NeT and IceCube can be entirely eliminated, because the induced number of UHE through-going and starting events in IceCube are predicted to be negligible.
We argued that this conclusion is robust against varying the energy spectrum, time duration (from minutes to months) and the elevation of the incoming flux.

We have shown that both transient flares of  AGNs and choked GRB sources  at the distances of $\sim $Gpc can be suitable sources for the dark flux because of two reasons: (1) Their enormous energy release can account for the energy of the $X$ flux; (2) The magnetic field inside them can convert active neutrinos to the $X$ particles which produce the signal at Earth. The column density at the source should be larger than 200 ${\rm t/cm}^2$ to ``choke" both the produced gamma ray and the UHE active neutrino flux. Such a column density is achievable  in a ``choked" GRB source \cite{Meszaros:2001ms}.

Our model includes a pair of fermions $X$ and $X'$ which are neutral  under the SM interactions, with a portal of light gauge, $Z'$, to the matter fields. We show that for $m_{Z'}\sim 300$~MeV, the coherent scattering off nuclei dominate over the deep inelastic as well as the quasi-elastic scatterings. $Z'$ can be identified as the gauge boson of  $B-3L_\mu$  or the dark photon. We have taken the couplings of $Z'$ to the baryons  just below the present bound $(\sim 10^{-4})$ which means $Z'$ is expected to show up at the experiments optimal for searching for light vector boson: For $B-3L_\mu$ gauge boson, the neutrino oscillation  and  Coherent Elastic $\nu$ Nucleus Scattering (CE$\nu$NS) experiments \cite{Coloma:2023ixt,AtzoriCorona:2022moj} as well as setups such as 
SHIFT@LHC \cite{Niedziela:2024khw} are the promising experiments to discover $Z'$. For the case of  dark photon, the entire parameter space of relevance to the KM3-230213A event can be probed by LHCb  \cite{LHCb:2019vmc}. 

The $X$ and $X'$ pair can  be produced in the $pp$ collision experiments via their coupling to $Z'$.  Then, the signal will be a dimuon pair with an invariant mass at $m_\pi$ from $X' \to \phi X$ and then $\phi \to \mu \bar{\mu}$ with a displaced vertex of $\sim 10^{-2}{\rm cm}(E_\mu+E_{\bar{\mu}})/(0.1~{\rm GeV})$. 
During the  run 3 LHC (during high luminosity LHC), we expect about $6\times 10^4$ signal events ($6\times 10^5$ signal events). The main detectors of LHC can resolve such a displaced vertex.  It is however beyond the scope of this paper to evaluate the discovery potential, considering the realistic background and the effects of the cuts and the efficiencies of the LHC.

We have also discussed the cosmological implications of our model.  All new particles except $X$ and $\nu_R$ decay well before the QCD transition era so they cannot affect the cosmological observables. The $X\bar{X}$ pair can annihilate to a $Z'$ pair with a cross section larger than 1 pb so the $X$ relic density cannot overclose the universe. $X$ can be considered a subdominant dark matter component but with a suppressed contribution of $\Omega_{X}/\Omega_{DM}\lsim 10^{-5}$. $\nu_R$ can contribute to the extra relativistic degrees of freedom. Since $\nu_R$ decouples from the plasma before the QCD phase transition, its contribution to the effective number of neutrinos is suppressed to 0.17 by the entropy pump to the plasma.  Such a value of $\Delta N_{eff}$ can be tolerated within the present bound \cite{Planck:2018vyg} but can be tested by future improvements on the cosmological data.

\begin{acknowledgments}
We thank Bhupal Dev, Bhaskar Dutta, Aparajitha Karthikeyan, Writasree Maitra, Louis Strigari and Ankur Verma for discussions and for coordinating the submission of our papers~\cite{Dev:2025czz}.
We also thank Vedran Brdar, Dibya  Chattopadhyay, Gustavo Alves, Ting Cheng, Pedro Machado, Maxim Pospelov, and Thomas Schwemberger for discussions on this topic.
The work of MH is supported by the Neutrino Theory Network Program Grant \#DE-AC02-07CHI11359 and the US DOE Award \#DE-SC0020250. This project has received funding from the European Union’s Horizon Europe research and innovation programme under the Marie Skłodowska-Curie Staff Exchange grant agreement No 101086085 – ASYMMETRY.
\end{acknowledgments}

\bibliographystyle{apsrev4-1}
\bibliography{main}{}

\end{document}